\documentclass{article}

\usepackage[preprint]{airov24}
\usepackage{graphicx}
\usepackage{cleveref} 
\usepackage{picinpar}
\usepackage{multirow}
\usepackage{makecell}

\usepackage[utf8]{inputenc} 
\usepackage[T1]{fontenc}    
\usepackage{booktabs}       
\usepackage{amsfonts}       
\usepackage{nicefrac}       
\usepackage{microtype}      
\usepackage{xcolor}         
\usepackage{mdframed}

\newmdtheoremenv{theo}{Definition}

\title{AuditMAI: Towards An Infrastructure for \\  Continuous AI Auditing}

\author{%
  Laura Waltersdorfer \\
 TU Wien \& WU Wien\\
  \texttt{laura.waltersdorfer@wu.ac.at} \\
  \And
   Fajar J. Ekaputra \\
 WU Wien \& TU Wien\\
  \texttt{fajar.ekaputra@wu.ac.at} \\
  \And
    Tomasz Miksa \\
 TU Wien \& SBA Research\\
  \texttt{tomasz.miksa@sba-research.at} \\
  \And
    Marta Sabou \\
 WU Wien\\
  \texttt{marta.sabou@wu.ac.at} \\
}

\begin{document}

\maketitle

\begin{abstract}
Artificial Intelligence (AI) Auditability is a core requirement for achieving responsible AI system design. However, it is not yet a prominent design feature in current applications. 
%
Existing AI auditing tools typically lack integration features and remain as isolated approaches. This results in manual, high-effort, and mostly one-off AI audits, necessitating alternative methods.
Inspired by other domains such as finance, \textit{continuous AI auditing} is a promising direction to conduct regular assessments of AI systems. 
The issue remains, however, since the methods for continuous AI auditing are not mature yet at the moment.
%
%
To address this gap, we propose the \textit{Auditability Method for AI (AuditMAI)},
which is intended as a blueprint for an infrastructure towards continuous AI auditing.
For this purpose, we first clarified the definition of AI auditability based on literature. 
Secondly, we derived requirements from two industrial use cases for continuous AI auditing tool support.
Finally, we developed AuditMAI and discussed its elements as a blueprint for a continuous AI auditability infrastructure.
%
\end{abstract}

\section{Introduction}\label{sec:intro}
Due to the increasing reports of incidents caused by Artificial Intelligence (AI) systems~\cite{McGregor2020}, calls for countermeasures and alternative design principles for AI systems have been raised. 
AI auditability is often mentioned as one of the high-level requirements for responsible AI design \cite{ai2019high, unesco2021recommendation}, since AI audits are considered as relevant means to check and assess AI systems \cite{prem2023ethical,raji2020closing,bandy2021problematic}. 

Furthermore, a move towards \textit{continuous AI auditing} practices to empower regular assessments of critical AI systems is needed due to the expected increased deployments of AI systems in critical and sensitive domains~\cite{minkkinen2022continuous}.
\textit{Continuous AI auditing} --inspired by financial auditing-- implies automatic periodic control and risk assessments of AI systems.
However, the absence of clear standards, requirements, and technical tools leaves much space for ambiguity of how to achieve AI auditability especially continuous AI auditing.

We argue that current approaches in AI auditing are hampered by the following two challenges:

(C1) \textit{Unclear definitions of AI auditability and requirements}. 
Current AI auditing approaches are described in heterogeneous vocabulary (e.g., impact assessment, technical or governance audits), leading to different expectations and requirements \cite{ayling2022putting}. 
Although AI auditability is frequently mentioned as a desired system characteristic to build responsible algorithmic systems \cite{williams2022transparency}, existing AI Principles failed to provide concrete recommendations and requirements on how to implement AI auditability \cite{shneiderman2020bridging}.

(C2) \textit{Lack of Automated AI auditing approaches and methods}. 
AI auditing methods have been proposed and are mostly in the form of multiple one-time audits by external parties being conducted with detective purposes \cite{bandy2021problematic}.
Furthermore, the majority of published governance processes focus on manual trace and data collection \cite{ayling2022putting, minkkinen2022continuous}.
These facts reinforced the notion that continuous AI auditing practices \cite{minkkinen2022continuous,ojewale2024towards} are not yet established.
Audit artefacts are often described as static documents or plain text to be extracted and analysed by humans. Logging services are often only tailored to specific components and provide limited to no flexible configuration. 
Only a few approaches provide ways to increase the degree of automation through technical AI audits to increase the repeatability of audits and, thus, their frequency \cite{minkkinen2022continuous, ojewale2024towards}. 

While human oversight, evaluation, and interpretation --in the form of manual efforts-- is and will be of crucial importance in AI auditability, increasing the degree of automation through continuous AI auditing is crucial towards achieving responsible AI design~\cite{ai2019high, unesco2021recommendation}.
Therefore, to overcome these challenges, in this paper, we focus on the following research question:

\begin{center}
\textit{What are key elements of an infrastructure to enable continuous AI auditability?}
\end{center}



To answer this question, we propose \textit{AuditMAI (Auditability Method for AI)}, a framework to enable continuous AI audits.
We describe key elements of an infrastructure enabling continuous AI auditing based on three levels: 1) knowledge, 2) process, and 3) architecture. 
To frame the scope of AuditMAI, we analysed literature for a working definition of AI auditability.
In contrast to the top-down approach for the definition, we derived four main requirements bottom-up from two industrial research projects \cite{Ekaputra2021,breit2023combining} focused on enabling continuous AI auditing. 
AuditMAI is comprised of three views: Knowledge, process, and architecture, targeting different interaction levels.
AuditMAI framework aims to close the gap in current auditability research by offering a holistic view of essential elements relevant for establishing continuous AI auditing.

The rest of this paper is structured as follows: In \Cref{sec:def}, we discuss related work and propose our own AI auditability definition. 
In \Cref{sec:uc}, we present our two industrial use cases and derive requirements (cf. \Cref{fig:requirements}) for technical support of continuous AI auditing.
In \Cref{sec:auditmai}, we propose AuditMAI (cf. \Cref{fig:method}), distilling our results and finally concluding in \Cref{sec:conclusion}. 

\section{AI Auditability Definition}
\label{sec:def}
AI auditability is frequently mentioned in responsible AI principles \cite{diaz2023connecting}.
However, despite the common description of this concept, a well-accepted definition with concrete technical implementation is missing. 
This lack of concrete requirements and definitions is hampering the creation of a continuous AI auditing framework. 
To overcome this gap, we summarized common themes in AI auditability (\Cref{sec:literature}) as a foundation for our AI auditability definition to scope our context (\Cref{sec:own_def}).

\subsection{Background: Common Discussion Themes in AI Auditability}\label{sec:literature}
In this section, we briefly discuss the following common themes on AI auditability: (i) AI audit goals, (ii) AI audit techniques, (iii) AI audit methods, (iv) AI auditing tools and (v) challenges for AI auditability.

\textit{Goals.} AI auditability is often aimed to achieve accountability \cite{cobbe2021reviewable, diaz2023connecting, cloete2021auditable} and transparency (either as a goal \cite{cobbe2021reviewable,cloete2021auditable} or enabler of AI auditability \cite{diaz2023connecting}).
Accountability is the obligation of providing information related to standardization and regulation or consequences after incidents, while transparency means the provision of relevant information about the system.

\textit{Techniques.} Audit traces are mentioned to be essential to AI auditability \cite{Brundage2020,shneiderman2020bridging}, meaning record and log-keeping support provision of required information for AI auditability \cite{cobbe2021reviewable, cloete2021auditable}.
Traces are collected to provide a holistic overview, calling for inspection of the entire system context:
Such information includes technical aspects such as data and algorithms and also contextual information or design decisions \cite{cobbe2021reviewable, ai2019high, diaz2023connecting}, but what exactly is left to be decided at the practical level.

\textit{Methods.} The following methods do not specifically define AI auditability but rather provide methods for different types of AI audits: Internal AI Audit methods~\cite{raji2020closing}, specific model types such as Large Language Models \cite{mokander2023auditing} or for specific purposes, e.g. ethics \cite{mokander2023operationalising}.
While these methods provide valuable insights, they do not focus much on technical aspects of audit data models and tool support. 

\textit{Tools.} 
A multitude of tools have been developed due to rising interest in responsible AI: Surveys collect frameworks and tools for ethical AI audits \cite{ayling2022putting}, continuous AI auditing \cite{minkkinen2022continuous}, and a broader context \cite{ojewale2024towards}.
While these attempts establish a useful basis by categorizing single tools, there are remaining challenges: 
i) \textit{tools} is a fuzzy term in AI auditing, not necessarily describing executable software but also guidelines \cite{ayling2022putting}, 
ii) isolated approaches since often tools are created for specific application contexts and use cases, it is unclear how to combine them along an AI auditing workflow \cite{ojewale2024towards}, and 
iii) context-dependence \cite{wfp2023}, e.g., the World Privacy Forum has identified that many popular fairness auditing tools implement US-specific legislation (\textit{US Four-Fifths Employment Rule for AI Fairness}) \cite{wfp2023}, not suitable for the rest of the world.


\textit{Needs.} The need for a holistic assessment poses challenges to AI system providers and operators, because tracking everything is seldom feasible and also not sensible due to opacity  \cite{cobbe2021reviewable} or other issues (e.g. privacy, data management) \cite{cloete2021auditable}.
This makes the selection and access of critical traces essential.
Furthermore, traces typically come from very different system components in heterogeneous formats, so logs need to be integrated at a meaningful semantic level to gain insights from them.

\subsection{Proposed AI Auditability Definition}
\label{sec:own_def}
%
Following our discussion on the common themes in AI Auditability and to establish the basis for our work, we define AI auditability as the following: 

\begin{theo}
\label{def:auditability}
\textbf{AI auditability} is the ability of an auditor to obtain accurate and relevant auditable artefacts for answering concrete audit questions, when examining an AI system.
\end{theo}

\begin{figure}
    \centering
    \includegraphics[scale=0.75]{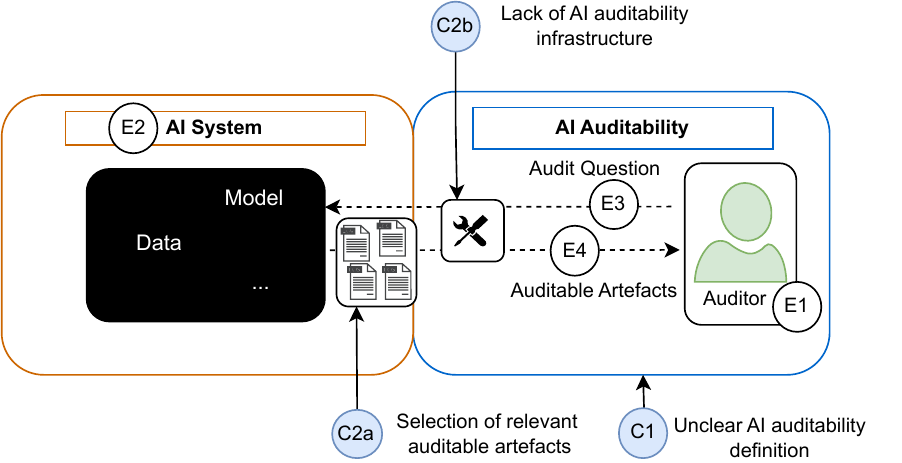}
    \caption{Auditability Definition Key Elements (E1-E4) and Challenges (C1-C2b)}
    \label{fig:def}
\end{figure}

To further clarify our definition, we explain the key AI auditability elements mentioned within our definition (\textit{auditor}, \textit{AI system}, \textit{audit questions} and \textit{auditable artefacts}) in the following text and illustrate them in \Cref{fig:def} (E1-E4) together with two identified challenges (C2a and C2b). 


     \textit{(1) Auditor} is any stakeholder being authorized to examine a given AI system for a specific purpose, typically by asking a set of audit questions. In literature, auditors are classified as being first, second and third party auditors, as well as internal and external auditors \cite{raji2022outsider}.

     \textit{(2) AI system} includes its algorithmic and technical components, as well as non-technical components, e.g. implicit knowledge involving humans or organisational aspects. However, the major challenge remains for system developers and providers what exactly needs to be captured. 
     
     \textit{(3) Audit questions} are defining the information needs for auditing purposes. These questions can be related to applicable legislation or certification standards.
    Some questions target specific AI system components (e.g., machine learning component) or the entire AI system (e.g., overall performance statistics) and have different audit objectives (e.g., fairness or robustness). 
    An example audit question is \textit{Who is the creator of this AI model?}, and a more complicated one is \textit{Which bias mitigation measures were taken throughout the development of this AI model?}
    Other forms of audit questions might be test cases to mitigate AI risks related to the system domain or task and also the chosen AI model (e.g., testing for adversarial input).
     
    \textit{(4) Auditable artefacts} 
    (also including audit traces, data statements, logs \cite{Brundage2020,shneiderman2020bridging,weigand2013conceptualizing,cloete2021auditable}), are the evidence used during AI audits to verify and validate the core functionality of an AI system by answering audit questions.
    We distinguish between \textit{static} and \textit{dynamic} artefacts:
    \textit{Static} artefacts include documentation that is less frequently updated, e.g. design decisions. Raji et al. \cite{raji2019actionable} describe examples for more static auditable artefacts for internal audits, e.g. stakeholder map. 
    \textit{Dynamic} artefacts, including \textbf{audit traces} collected from the live systems, from the training or execution phase of a system (e.g. status or output logs).
    Furthermore, to avoid opacity and other issues as mentioned in \Cref{sec:literature}, not all information that can be collected should be collected. 
    Thus, it is essential to consider \textit{relevancy} and \textit{accuracy} of auditable artefacts. Both factors are closely linked to the overall audit goal and thus, related audit questions. 




The definition in this section clarified and addressed the first challenge (C1) \textit{unclear AI auditability definiton} mentioned in~\Cref{sec:intro}. Furthermore, it also helped to clarify the scope of the second challenge (C2) \textit{lack of automated AI auditing approaches and methods}.
This challenge consists now of 
(C2a) the selection of relevant traces and documents within the AI system context, and 
(C2b) the provision of an AI auditability infrastructure to support the continuous AI auditing process.


\section{Use-case based AI Auditability Requirements}
\label{sec:usecases}

We present two use cases (UC) from industrial research projects to derive requirements for the development of continuous AI auditing tool support in order to address challenge C2 (cf.~\Cref{sec:intro}). 
To this end, we first discuss them based on the previously introduced AI auditability elements, namely, \textit{Auditors}, \textit{AI system}, \textit{Audit questions} and
\textit{Auditable artefacts} (cf. \Cref{sec:own_def}).
Then, we summarize requirements derived from the use cases (cf. Fig. \ref{fig:requirements} and \Cref{tab:uc}).

\label{sec:uc}
\subsection{Use Case Analysis}
The use cases are from different domains and UC partners: UC1 is situated in the environmental-legal domain in cooperation with a federal agency~\cite{breit2023combining}; while UC2 is conducted in the medical domain with several small companies~\cite{Ekaputra2021}.

\textbf{UC 1}
The use case covers the ecological and legal domain of a federal agency dealing with environmental permits.
The to-be audited AI system is to complement human manual extraction from legal permits through automatic extraction through machine learning.
Key information items are extracted from legal permits (PDFs) to add metadata information of those permits for search purposes (e.g. operator or applicable law). 

\textit{Auditors.} 
\textit{System operators}, who are no machine learning experts are interested in underlying patterns of wrong predictions and frequently corrected key information types.
\textit{Legal analysts} are interested in single executions. Both qualify as internal auditors.

\textit{AI System.} Main components are: The user interface for document upload and transformation of PDF documents in machine-readable format.
After data extraction from the documents, four services (2 ML and 2 non-ML services and an ontology) provide suggestions for key entity types.
Finally, users can correct suggestions by the AI system.

\textit{Audit Question.} Examples included: \textit{What is the average confidence value per entity type over time?} to analyse the overall model fit continuously.
Another audit question was: \textit{Did this system execution run successfully?} to ensure the correct functioning of single system runs.

\textit{Auditable Artefact.} Traces from the identified main AI system components need to be collected, such as version numbers, timestamps, and completion states, but also more concrete ones, such as the confidence values for key entity types or correction states. 





\textbf{UC 2}
This use case is situated in the medical domain of startups integrating health data from multiple application providers, under regulation, e.g. the General Data Protection Regulation (GDPR) in the European Union.
Small health application providers choose for compliance reasons to store their users' data in a cloud platform.
This way, they do not have to develop and run the platform themselves.
The centralised platform manages data access to permit data analysis through privacy-preserving data analysis.
Service provider upload their data with the applicable consent rights from users to the platform.
Researchers can specify study criteria for applicable health data. Consent management is managed by the platform.

\textit{Auditors.} \textit{Analysts} are interested in which software libraries were used for a certain analysis.
\textit{System operators} can inquire about the integrated logging concerning the data access and analysis actions performed by analysts for accountability of past studies. The former are considered external, while the latter are internal auditors.

\textit{AI System.} Components include: user data upload, data selection (including consent management) and analysis (through ML models). 

\textit{Audit Question.} Example questions are: \textit{Was the consent evaluation executed
during data collection} to ensure compliance. To audit data analysis: \textit{Which software libraries have been used to analyse data in a specific study?}

\textit{Auditable Artefact.} Traces from many sources needed to be collected. For example, the consent management (in \texttt{RDF}) for the consent result and the analysis module (in \texttt{R}) for used libraries.

\subsection{Auditability Requirements} 
\label{sec:uc-requirements}

\begin{figure}
    \centering
    \includegraphics[width=\textwidth]{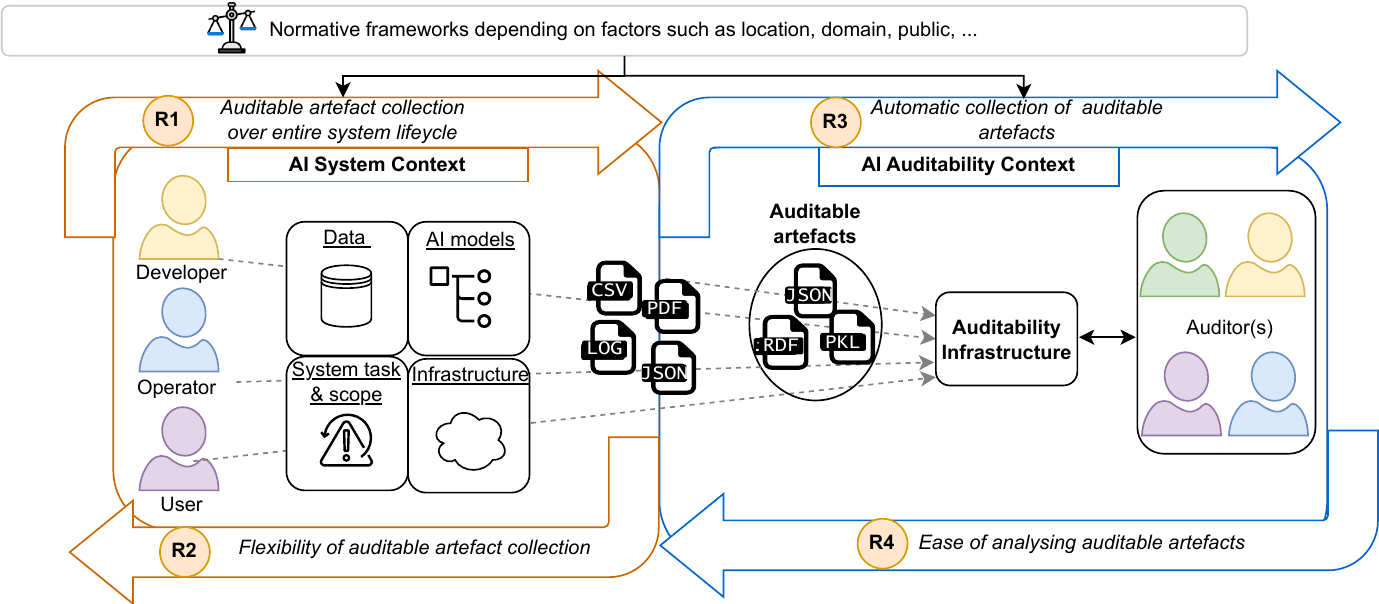}
    \caption{A schematic overview of the UCs and derived requirements (R1-R4)}
    \label{fig:requirements}
\end{figure}

Several requirements can be derived from the generic continuous auditing approach~\cite{minkkinen2022continuous}. However, since we aim to specify more concrete requirements specifically for AI systems, we derived a set of requirements through multiple rounds of workshops with the project partners to design and implement tool support for continuous AI auditing. We visualised the requirements (R1-R4) in \Cref{fig:requirements} and explain them in the following.

R1: \textit{Identification of relevant auditable artefacts logging over different AI lifecycle phases}: Similar to software engineering, AI systems undergo lifecycle phases from design until decommissioning. A key challenge is the identification of relevant audit information.
In both UCs, the audit questions were iteratively derived and refined. In UC 1, the audit focus was put on the exploitation phase capturing runtime phrases of the system, with just a limited number of training parameters being captured, while in UC 2, the focus was on the overall system, also tracking main data flows during execution.
Supporting the identification of relevant artefacts is vital to the preparation of AI auditing.

R2: \textit{Flexibility of auditable artefact collection mechanisms}: AI systems can handle multi-modal data ranging from text, images, video or sound, encoded in various formats. 
In both UCs, we encountered different software languages (\texttt{Python, R, Java}) and libraries, as well as artefact formats, such as \texttt{JSON}, \texttt{CSV}, \texttt{TIFF}, or \texttt{RDF}.
Some log structures were predefined, while others could be adapted to the AI auditing needs.
Flexibility is also needed in terms of granularity, since for different auditing objectives different information is needed (e.g. fairness vs. accountability).
This flexibility in the system context must also be represented and feasible for AI auditability by offering flexible and variable collection mechanisms.

R3: \textit{Automation of auditable artefact collection, transformation and management}: Auditable artefacts are created by multiple components, in different system states as described in the previous requirements and include various artefacts in the design phase or high numbers of transactions per system execution during system execution.
Therefore, collection, transformation and management of these artefacts during the system development and operation is critical, while manual logging is undesired.
For both UCs, automatic collection and management was needed for effective AI auditability. 
Integrating heterogeneous trace data into a metadata representation is also needed to provide improved analysis capabilities.

R4: \textit{Ease of analysing auditable artefacts}: The artefacts need to be analysed to obtain audit results and plan further steps based on the outcome. Analysis capabilities include the investigation of auditable artefacts, writing of queries or computing metrics and visualisation of results for communication or reporting purposes. Different levels of expertise and professional backgrounds of potential auditors require easy analytics and reporting possibilities.  
In UC 1, there was only limited technical knowledge to be expected from the internal auditors and for UC 2, main requirement was to offer a query interface with pre-defined queries and possibility to write custom queries.

\begin{table}
    \centering
    \begin{tabular}{|p{2.2cm}|p{5.3cm}|p{5.3cm}|}
                  \hline

     \textit{Elements} &  \textit{UC 1} & \textit{UC 2} \\
       \toprule
           \textbf{Auditor}  & (both internal) System operators, legal analysts & (internal) System operators, (external) analysts \\ \hline
       \textbf{AI System}  & UI, 2 ML components, 2 non-ML services, ontology & UI, consent check, ML-based analysis \\ \hline
         \textbf{AQ System Operator}  & What is the average confidence value per key entity type? &Was the consent evaluation executed during data collection? \\ \hline 
           \textbf{AQ Analyst}  & Did this system execution
run successfully?& Which software libraries have been used in a specific study?\\ \hline
      \textbf{Auditable Artefacts}    & RDF traces, (ML) services output, user corrections & R scripts, consent result\\ \hline
       \textbf{Requirements}    & R1 (focus on exploitation phase of ML systems), R2 (log structure was partly pre-defined), R3 (full automation), R4 (query interface with pre-defined queries, detailed custom queries) & R1 (focus on data flows in entire system), R2 (log structure was flexible), R3 (full automation), R4 (query interface with pre-defined queries, custom ones) \\ \hline
    \end{tabular}
    \caption{AI Auditability elements in UCs - AQ (Audit Question)}
    \label{tab:uc}
\end{table}

All of these requirements, focusing on the handling and management of auditable artefacts are enablers for AI auditability.
To the best of our knowledge, there does not yet exist a tool suite to cover all these requirements, but selected aspects.
For example, MLOps tools e.g. \textit{MLflow} provide capabilities especially for R2 and R3. Logging and analysis of common parameters and metadata for ML models can be easily executed with such tools. The selection and higher degree of granularity for specific parameters and metrics are more difficult to achieve.
For R1 especially auditable artefacts beyond common MLOps artefacts are not covered.
Also for R4 the integration and analysis mechanisms of auditable artefacts is not supported beyond common ones.
To address this gap, we propose AuditMAI.

\section{AuditMAI Framework}
\label{sec:auditmai}

To address both the identified challenges (cf. \Cref{sec:own_def}: C2a-C2b) as well as the requirements (cf. \Cref{sec:uc-requirements}: R1-R4), we present \textbf{AuditMAI} (Auditability Method for AI) framework.
The aim is to provide a blueprint for AI auditability through technical tools for the auditing process.
The framework was iteratively built  and refined based on our experiences of past projects. AuditMAI is partly implemented through our prototype AuditBox, leveraging semantic web technologies for auditable artefact integration and management \cite{Ekaputra2021,breit2023combining}. 
\paragraph{Framework Description.}
AuditMAI is divided into three views targeting different levels of the AI auditability infrastructure: (i) Knowledge View, (ii) Process View and (iii) Architecture View.
The framework integrates the aspects of knowledge management, process understanding and architectural design for supporting continuous AI auditing showing relevant flows between each layer. In the following, we discuss them:


\textbf{I. Knowledge View}  - gathers knowledge about derived AI audit questions, test cases and risks, documentation standards and existing tools and metrics. It consists of three knowledge bases:

\textit{Audit Questions}, (cf. \Cref{sec:def})
 suggest which information needs to be collected during the audit process.
Initial collections have been made for different purposes e.g. explainability \cite{liao2020questioning} or accountability \cite{naja2022using} or trustworthy AI \cite{poretschkin2021leitfaden}. 
Initial risk taxonomies have been synthesised \cite{raji2022fallacy} through AI Incident databases, such as Incident DB \cite{McGregor2020} and AIAAIC\footnote{\url{https://www.aiaaic.org}}. 
  
    \textit{Documentation Standards}, collect relevant metadata standards to describe and document relevant information of auditable artefacts. 
Well-known approaches such as ModelCards \cite{mitchell2019model} and Datasheets \cite{gebru2021datasheets} have reached a certain adoption due their integration on popular online platforms e.g. HuggingFace or Github. However, current research indicates that AI documentation is failing to adequately include and communicate relevant AI system features \cite{bhat2023aspirations,longpre2023data}.

\textit{Tool and Metrics} differ depending on task and domain. Knowledge bases such as the OECD Catalogue\footnote{ \url{https://oecd.ai/en/catalogue/overview}} or surveys \cite{minkkinen2022continuous, ayling2022putting} are helpful to investigate available tools.
However, since terminology in AI auditing is still fuzzy, attempts such as the recently published AI audit tooling landscape map\footnote{ \url{https://tools.auditing-ai.com}} aim to go beyond prior classifications by providing a more comprehensive audit tool taxonomy.
The same is needed to organise metrics and evaluation techniques for AI auditability.

\textbf{II. Process View} - focuses on the auditing process steps incorporating \textit{audit questions} and \textit{auditable artefacts}. It consists of four auditing steps:
\begin{figure}
    \centering    \includegraphics[width=\textwidth]{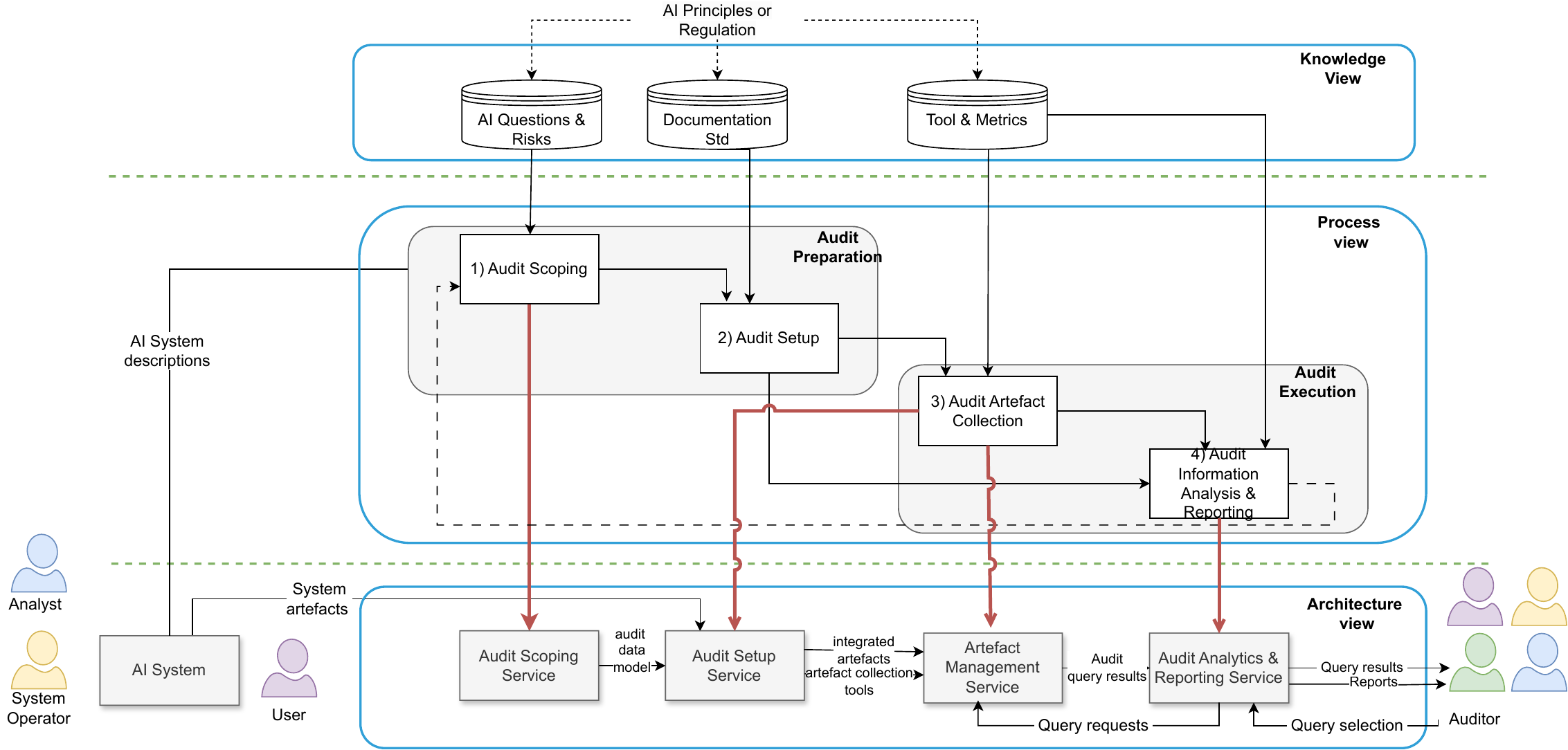}
    \caption{AuditMAI: Combining knowledge management, process understanding and architectural design for AI Auditability Infrastructure}
    \label{fig:method}
\end{figure}

\textit{Step 1: Audit Scoping.} To scope the audit process, two inputs are needed: i) the AI system description describing main data flows and algorithmic modules and ii) the overall audit goal, e.g. transparency or fairness.
Both descriptions are then also input to
the architectural component (cf. \textit{Audit Scoping Service}) to provide recommendations and subsequent selection of applicable audit questions and risks also based on the input from the \textit{Knowledge View}.
     
\textit{Step 2: Audit Setup.} After scoping the audit, this step is concerned with setting up the collection of auditable artefacts.
These artefacts are derived based on the AI system description from the previous step.
Furthermore, identified artefacts need to be integrated into a common view across the entire AI system to answer complex audit questions e.g. in a graph-based representation (cf. architectural component \textit{Audit Setup Service}).


\textit{Step 3: Audit Artefact Collection.} This step deals with collecting \textit{auditable artefacts} during the runtime of a system.
Ideally, artefact collection is automated, with only limited data being collected or tested manually (cf. architectural component \textit{Audit Collection Service)}.

\textit{Step 4: Audit Analysis and Reporting.}
     In the final step, \textit{auditable artefacts} are analysed to report results of the audit process to stakeholders and identify system improvements (cf. architectural component \textit{Audit Analytics \& Reporting Service}).
     Another audit process can be started then.

\textbf{III. Architecture View} - describes technical services to support the continuous AI auditing process, especially the definition of \textit{audit questions} and management of relevant \textit{auditable artefacts}. The four requirements (cf. \Cref{sec:uc-requirements}) are targeted by the following four services:  

\textit{Audit Scoping Service.} The audit scoping includes the selection of applicable audit questions, risks and test cases (R1 \textit{Identification of relevant auditable artefacts)}.
This is highly dependent on the AI system, context and domain.
The AI system workflow description with main data flows and algorithmic components assists in suggesting example audit questions or test cases in each phase of the system lifecycle. In line with continuous auditing, knowledge from past experiences is integrated, e.g., from AI incident databases, and an audit data model is output for subsequent steps.

\textit{Audit Setup Service.} This service aims to prepare the collection of auditable artefacts from available data sources (\textit{R2 Flexibility of auditable artefact collection mechanisms}). Example data sources can be textual descriptions (e.g., Model Cards \cite{mitchell2019model}), automatic MLOps tools (e.g., MLflow\footnote{ \url{https://mlflow.org}}), or specific automatic ones (e.g., Aequitas for fairness \cite{saleiro2018aequitas}).
Also, logs might be already collected for other purposes (e.g., debugging or security) and could be extended for auditing purposes.
An overview of the required auditable artefacts and the needed artefact collection tools is the output of this step. 

\textit{Artefact Management Service.} This service is responsible for storing and persisting collected artefacts (\textit{R3 Automation of auditable artefact management}) for analysis and reporting. The choice of storage and retrieval of artefacts might depend on organisational contexts and requirements, such as given infrastructure.
Due to the heterogeneity of auditable artefacts and to answer complex audit queries, data integration is important e.g. through Knowledge Graphs and Semantic Web technologies \cite{naja2022using,breit2023combining}. 

\textit{Audit Artefact Analytics and Reporting Service.} This component provides analysis and investigation capabilities of auditable artefacts.
Artefacts can be analysed through a provided dashboard, predefined or custom queries or generated reports accessing the artefact management service (\textit{R4 Ease of analysing auditable artefacts}).

\section{Conclusion}
\label{sec:conclusion}

Through AuditMAI, we provide a framework for AI auditability infrastructure for continuous AI auditing.
We aim to counteract unclear requirements hampering both process and technical maturity in AI auditability and the currently fragmented emerging AI audit tools landscape. 
To achieve this, the framework is scoped around our AI auditability definition and requirements from industrial projects structured into three different views.  
The aim is to reduce manual efforts and support preventive and continuous AI auditing.


\textbf{Discussion.} 
While the use cases provided important insights into the needs and requirements for continuous AI auditing practices, they might be subject to selection and observer bias. 

\textbf{Future Work.} We plan to further analyze the capabilities of existing AI auditing tools and audit data models to extend our work on the AuditBox tool for AI Auditability \cite{breit2023combining}, specifically to cover all four steps of the AuditMAI framework.
Furthermore, we plan to conduct a comprehensive evaluation of the AuditMAI framework with AuditBox as the reference implementation.
%

\bibliographystyle{unsrtnat}

\bibliography{references}

\begin{thebibliography}{32}
\providecommand{\natexlab}[1]{#1}
\providecommand{\url}[1]{\texttt{#1}}
\expandafter\ifx\csname urlstyle\endcsname\relax
  \providecommand{\doi}[1]{doi: #1}\else
  \providecommand{\doi}{doi: \begingroup \urlstyle{rm}\Url}\fi

\bibitem[McGregor(2021)]{McGregor2020}
Sean McGregor.
\newblock {Preventing Repeated Real World AI Failures by Cataloging Incidents: The AI Incident Database}.
\newblock In \emph{Proceedings of the AAAI Conference on Artificial Intelligence}, 2021.

\bibitem[AI(2019)]{ai2019high}
HLEG AI.
\newblock \emph{{High-level Expert Group on Artificial Intelligence}}.
\newblock European Commission, 2019.

\bibitem[UNESCO(2021)]{unesco2021recommendation}
UNESCO.
\newblock {Recommendation on the Ethics of Artificial Intelligence}, 2021.

\bibitem[Prem(2023)]{prem2023ethical}
Erich Prem.
\newblock {From Ethical AI Frameworks to Tools: A Review of Approaches}.
\newblock \emph{AI and Ethics}, pages 1--18, 2023.

\bibitem[Raji et~al.(2020)Raji, Smart, White, Mitchell, Gebru, Hutchinson, Smith-Loud, Theron, and Barnes]{raji2020closing}
Inioluwa~Deborah Raji, Andrew Smart, Rebecca~N White, Margaret Mitchell, Timnit Gebru, Ben Hutchinson, Jamila Smith-Loud, Daniel Theron, and Parker Barnes.
\newblock {Closing the AI Acountability Gap: Defining an End-to-End Framework for Internal Algorithmic Auditing}.
\newblock In \emph{Proceedings of the 2020 Conference on Fairness, Accountability, and Transparency}, pages 33--44, 2020.

\bibitem[Bandy(2021)]{bandy2021problematic}
Jack Bandy.
\newblock {Problematic Machine Behavior: A Systematic Literature Review of Algorithm Audits}.
\newblock \emph{Proceedings of the ACM on Human-Computer Interaction}, 5\penalty0 (CSCW1):\penalty0 1--34, 2021.

\bibitem[Minkkinen et~al.(2022)Minkkinen, Laine, and M{\"a}ntym{\"a}ki]{minkkinen2022continuous}
Matti Minkkinen, Joakim Laine, and Matti M{\"a}ntym{\"a}ki.
\newblock {Continuous Auditing of Artificial Intelligence: A Conceptualization and Assessment of Tools and Frameworks}.
\newblock \emph{Digital Society}, 1\penalty0 (3):\penalty0 21, 2022.

\bibitem[Ayling and Chapman(2022)]{ayling2022putting}
Jacqui Ayling and Adriane Chapman.
\newblock {Putting AI Ethics to Work: Are the Tools Fit for Purpose?}
\newblock \emph{AI and Ethics}, 2\penalty0 (3):\penalty0 405--429, 2022.

\bibitem[Williams et~al.(2022)Williams, Cloete, Cobbe, Cottrill, Edwards, Markovic, Naja, Ryan, Singh, and Pang]{williams2022transparency}
Rebecca Williams, Richard Cloete, Jennifer Cobbe, Caitlin Cottrill, Peter Edwards, Milan Markovic, Iman Naja, Frances Ryan, Jatinder Singh, and Wei Pang.
\newblock {From Transparency to Accountability of Intelligent Systems: Moving beyond Aspirations}.
\newblock \emph{Data \& Policy}, 4:\penalty0 e7, 2022.

\bibitem[Shneiderman(2020)]{shneiderman2020bridging}
Ben Shneiderman.
\newblock {Bridging the Gap Between Ethics and Practice: Guidelines for Reliable, Safe, and Trustworthy Human-centered AI systems}.
\newblock \emph{ACM Transactions on Interactive Intelligent Systems (TiiS)}, 10\penalty0 (4):\penalty0 1--31, 2020.

\bibitem[Ojewale et~al.(2024)Ojewale, Steed, Vecchione, Birhane, and Raji]{ojewale2024towards}
Victor Ojewale, Ryan Steed, Briana Vecchione, Abeba Birhane, and Inioluwa~Deborah Raji.
\newblock {Towards AI Accountability Infrastructure: Gaps and Opportunities in AI Audit Tooling}.
\newblock \emph{arXiv preprint arXiv:2402.17861}, 2024.

\bibitem[Ekaputra et~al.(2021)Ekaputra, Ekelhart, Mayer, Miksa, {S}ar\v{c}evi\'{c}, Tsepelakis, and Waltersdorfer]{Ekaputra2021}
Fajar~J. Ekaputra, Andreas Ekelhart, Rudolf Mayer, Tomasz Miksa, Tanja {S}ar\v{c}evi\'{c}, Sotiris Tsepelakis, and Laura Waltersdorfer.
\newblock {Semantic-enabled Architecture for Auditable Privacy-Preserving Data Analysis}.
\newblock \emph{Semantic Web Journal}, 2021.

\bibitem[Breit et~al.(2023)Breit, Waltersdorfer, Ekaputra, Karampatakis, Miksa, and K{\"a}fer]{breit2023combining}
Anna Breit, Laura Waltersdorfer, Fajar~J Ekaputra, Sotirios Karampatakis, Tomasz Miksa, and Gregor K{\"a}fer.
\newblock {Combining Semantic Web and Machine Learning for Auditable Legal Key Element Extraction}.
\newblock In \emph{European Semantic Web Conference}, pages 609--624. Springer, 2023.

\bibitem[D{\'\i}az-Rodr{\'\i}guez et~al.(2023)D{\'\i}az-Rodr{\'\i}guez, Del~Ser, Coeckelbergh, de~Prado, Herrera-Viedma, and Herrera]{diaz2023connecting}
Natalia D{\'\i}az-Rodr{\'\i}guez, Javier Del~Ser, Mark Coeckelbergh, Marcos~L{\'o}pez de~Prado, Enrique Herrera-Viedma, and Francisco Herrera.
\newblock {Connecting the Dots in Trustworthy Artificial Intelligence: From AI Principles, Ethics, and Key Requirements to Responsible AI Systems and Regulation}.
\newblock \emph{Information Fusion}, page 101896, 2023.

\bibitem[Cobbe et~al.(2021)Cobbe, Lee, and Singh]{cobbe2021reviewable}
Jennifer Cobbe, Michelle Seng~Ah Lee, and Jatinder Singh.
\newblock {Reviewable Automated Decision-Making: A Framework for Accountable Algorithmic Systems}.
\newblock In \emph{Proceedings of the 2021 ACM Conference on Fairness, Accountability, and Transparency}, pages 598--609, 2021.

\bibitem[Cloete et~al.(2021)Cloete, Norval, and Singh]{cloete2021auditable}
Richard Cloete, Chris Norval, and Jatinder Singh.
\newblock Auditable augmented/mixed/virtual reality: The practicalities of mobile system transparency.
\newblock \emph{Proceedings of the ACM on Interactive, Mobile, Wearable and Ubiquitous Technologies}, 5\penalty0 (4):\penalty0 1--24, 2021.

\bibitem[Brundage et~al.(2020)Brundage, Avin, Wang, Belfield, Krueger, Hadfield, Khlaaf, Yang, Toner, Fong, Maharaj, Koh, Hooker, Leung, Trask, Bluemke, Lebensold, O'Keefe, Koren, Ryffel, Rubinovitz, Besiroglu, Carugati, Clark, Eckersley, de~Haas, Johnson, Laurie, Ingerman, Krawczuk, Askell, Cammarota, Lohn, Krueger, Stix, Henderson, Graham, Prunkl, Martin, Seger, Zilberman, hEigeartaigh, Kroeger, Sastry, Kagan, Weller, Tse, Barnes, Dafoe, Scharre, Herbert-Voss, Rasser, Sodhani, Flynn, Gilbert, Dyer, Khan, Bengio, and Anderljung]{Brundage2020}
Miles Brundage, Shahar Avin, Jasmine Wang, Haydn Belfield, Gretchen Krueger, Gillian Hadfield, Heidy Khlaaf, Jingying Yang, Helen Toner, Ruth Fong, Tegan Maharaj, Pang~Wei Koh, Sara Hooker, Jade Leung, Andrew Trask, Emma Bluemke, Jonathan Lebensold, Cullen O'Keefe, Mark Koren, Théo Ryffel, JB~Rubinovitz, Tamay Besiroglu, Federica Carugati, Jack Clark, Peter Eckersley, Sarah de~Haas, Maritza Johnson, Ben Laurie, Alex Ingerman, Igor Krawczuk, Amanda Askell, Rosario Cammarota, Andrew Lohn, David Krueger, Charlotte Stix, Peter Henderson, Logan Graham, Carina Prunkl, Bianca Martin, Elizabeth Seger, Noa Zilberman, Sean~O hEigeartaigh, Frens Kroeger, Girish Sastry, Rebecca Kagan, Adrian Weller, Brian Tse, Elizabeth Barnes, Allan Dafoe, Paul Scharre, Ariel Herbert-Voss, Martijn Rasser, Shagun Sodhani, Carrick Flynn, Thomas~Krendl Gilbert, Lisa Dyer, Saif Khan, Yoshua Bengio, and Markus Anderljung.
\newblock {Toward Trustworthy AI Development: Mechanisms for Supporting Verifiable Claims}.
\newblock 4 2020.
\newblock URL \url{http://arxiv.org/abs/2004.07213}.

\bibitem[M{\"o}kander et~al.(2023)M{\"o}kander, Schuett, Kirk, and Floridi]{mokander2023auditing}
Jakob M{\"o}kander, Jonas Schuett, Hannah~Rose Kirk, and Luciano Floridi.
\newblock {Auditing Large Language Models: A Three-Layered Approach}.
\newblock \emph{AI and Ethics}, pages 1--31, 2023.

\bibitem[M{\"o}kander and Floridi(2023)]{mokander2023operationalising}
Jakob M{\"o}kander and Luciano Floridi.
\newblock {Operationalising AI Governance through Ethics-Based Auditing: an Industry Case Study}.
\newblock \emph{AI and Ethics}, 3\penalty0 (2):\penalty0 451--468, 2023.

\bibitem[Forum(2023)]{wfp2023}
World~Privacy Forum.
\newblock \emph{{Risky Analysis: Assessing and Improving AI Governance Tools}}.
\newblock 2023.

\bibitem[Raji et~al.(2022{\natexlab{a}})Raji, Xu, Honigsberg, and Ho]{raji2022outsider}
Inioluwa~Deborah Raji, Peggy Xu, Colleen Honigsberg, and Daniel Ho.
\newblock {Outsider Oversight: Designing a Third Party Audit Ecosystem for AI Governance}.
\newblock In \emph{Proceedings of the 2022 AAAI/ACM Conference on AI, Ethics, and Society}, pages 557--571, 2022{\natexlab{a}}.

\bibitem[Weigand et~al.(2013)Weigand, Johannesson, Andersson, Bergholtz, Bukhsh, Deneckere, and Proper]{weigand2013conceptualizing}
Hans Weigand, Paul Johannesson, Birger Andersson, Maria Bergholtz, Faiza~Allah Bukhsh, R~Deneckere, and H~Proper.
\newblock {Conceptualizing Auditability}.
\newblock In \emph{CAiSE forum}, pages 49--56, 2013.

\bibitem[Raji and Buolamwini(2019)]{raji2019actionable}
Inioluwa~Deborah Raji and Joy Buolamwini.
\newblock Actionable auditing: Investigating the impact of publicly naming biased performance results of commercial ai products.
\newblock In \emph{Proceedings of the 2019 AAAI/ACM Conference on AI, Ethics, and Society}, pages 429--435, 2019.

\bibitem[Liao et~al.(2020)Liao, Gruen, and Miller]{liao2020questioning}
Q~Vera Liao, Daniel Gruen, and Sarah Miller.
\newblock {Questioning the AI: Informing Design Practices for Explainable AI User Experiences}.
\newblock In \emph{Proceedings of the 2020 CHI Conference on Human Factors in Computing Systems}, pages 1--15, 2020.

\bibitem[Naja et~al.(2022)Naja, Markovic, Edwards, Pang, Cottrill, and Williams]{naja2022using}
Iman Naja, Milan Markovic, Peter Edwards, Wei Pang, Caitlin Cottrill, and Rebecca Williams.
\newblock {Using Knowledge Graphs to Unlock Practical Collection, Integration, and Audit of AI Accountability Information}.
\newblock \emph{IEEE Access}, 10:\penalty0 74383--74411, 2022.

\bibitem[Poretschkin et~al.(2021)Poretschkin, Schmitz, Akila, Adilova, Becker, Cremers, Hecker, Houben, Mock, Rosenzweig, et~al.]{poretschkin2021leitfaden}
Maximilian Poretschkin, Anna Schmitz, Maram Akila, Linara Adilova, Daniel Becker, Armin~B Cremers, Dirk Hecker, Sebastian Houben, Michael Mock, Julia Rosenzweig, et~al.
\newblock {Leitfaden zur Gestaltung vertrauensw{\"u}rdiger k{\"u}nstlicher Intelligenz}.
\newblock \emph{KI-Pr{\"u}fkatalog. Sankt Augustin: Fraunhofer-Institut f{\"u}r Intelligente Analyse-und Informationssysteme IAIS}, 2021.

\bibitem[Raji et~al.(2022{\natexlab{b}})Raji, Kumar, Horowitz, and Selbst]{raji2022fallacy}
Inioluwa~Deborah Raji, I~Elizabeth Kumar, Aaron Horowitz, and Andrew Selbst.
\newblock {The Fallacy of AI Functionality}.
\newblock In \emph{Proceedings of the 2022 ACM Conference on Fairness, Accountability, and Transparency}, pages 959--972, 2022{\natexlab{b}}.

\bibitem[Mitchell et~al.(2019)Mitchell, Wu, Zaldivar, Barnes, Vasserman, Hutchinson, Spitzer, Raji, and Gebru]{mitchell2019model}
Margaret Mitchell, Simone Wu, Andrew Zaldivar, Parker Barnes, Lucy Vasserman, Ben Hutchinson, Elena Spitzer, Inioluwa~Deborah Raji, and Timnit Gebru.
\newblock {Model Cards for Model Reporting}.
\newblock In \emph{Proceedings of the Conference on Fairness, Accountability, and Transparency}, pages 220--229, 2019.

\bibitem[Gebru et~al.(2021)Gebru, Morgenstern, Vecchione, Vaughan, Wallach, Iii, and Crawford]{gebru2021datasheets}
Timnit Gebru, Jamie Morgenstern, Briana Vecchione, Jennifer~Wortman Vaughan, Hanna Wallach, Hal~Daum{\'e} Iii, and Kate Crawford.
\newblock Datasheets for datasets.
\newblock \emph{Communications of the ACM}, 64\penalty0 (12):\penalty0 86--92, 2021.

\bibitem[Bhat et~al.(2023)Bhat, Coursey, Hu, Li, Nahar, Zhou, K{\"a}stner, and Guo]{bhat2023aspirations}
Avinash Bhat, Austin Coursey, Grace Hu, Sixian Li, Nadia Nahar, Shurui Zhou, Christian K{\"a}stner, and Jin~LC Guo.
\newblock {Aspirations and Practice of ML Model Documentation: Moving the Needle with Nudging and Traceability}.
\newblock In \emph{Proceedings of the 2023 CHI Conference on Human Factors in Computing Systems}, pages 1--17, 2023.

\bibitem[Longpre et~al.(2023)Longpre, Mahari, Chen, Obeng-Marnu, Sileo, Brannon, Muennighoff, Khazam, Kabbara, Perisetla, et~al.]{longpre2023data}
Shayne Longpre, Robert Mahari, Anthony Chen, Naana Obeng-Marnu, Damien Sileo, William Brannon, Niklas Muennighoff, Nathan Khazam, Jad Kabbara, Kartik Perisetla, et~al.
\newblock {The Data Provenance Initiative: A Large Scale Audit of Dataset Licensing \& Attribution in AI}.
\newblock \emph{arXiv preprint arXiv:2310.16787}, 2023.

\bibitem[Saleiro et~al.(2018)Saleiro, Kuester, Hinkson, London, Stevens, Anisfeld, Rodolfa, and Ghani]{saleiro2018aequitas}
Pedro Saleiro, Benedict Kuester, Loren Hinkson, Jesse London, Abby Stevens, Ari Anisfeld, Kit~T Rodolfa, and Rayid Ghani.
\newblock {Aequitas: A Bias and Fairness Audit Toolkit}.
\newblock \emph{arXiv preprint arXiv:1811.05577}, 2018.

\end{thebibliography}


\end{document}